\documentclass{article}

\usepackage{tabularx,booktabs} 
\usepackage{hyperref}
\usepackage{arxiv}
\usepackage[table,xcdraw]{xcolor}
\usepackage{lscape}
\usepackage[utf8]{inputenc} 
\usepackage[T1]{fontenc}    
\usepackage{hyperref}       
\usepackage{url}            
\usepackage{booktabs}       
\usepackage{amsfonts}       
\usepackage{nicefrac}       
\usepackage{microtype}      
\usepackage{lipsum}
\usepackage{graphicx} 
\usepackage{threeparttable, tablefootnote} 
\usepackage{array, makecell} 
\usepackage{booktabs} 
\usepackage[misc]{ifsym}
\graphicspath{ {images/} }

\usepackage{pifont}
\newcommand{\cmark}{\ding{51}}%

\usepackage{xcolor}

\definecolor{green(ryb)}{rgb}{0.4, 0.69, 0.2}

\usepackage[bf,font=footnotesize]{caption} 
\usepackage{enumitem}  
\usepackage{dirtytalk}

\usepackage{authblk}

\usepackage{titlesec}
\titlespacing\section{0pt}{12pt plus 4pt minus 2pt}{0pt plus 2pt minus 2pt}

\title{On the Composition and Limitations of Publicly Available COVID-19 X-Ray Imaging Datasets} 

\author[2,1,\Letter]{Beatriz~Garcia Santa Cruz}
\author[1]{Jan~Sölter}
\author[1]{Matias Nicolas~Bossa}
\author[1]{Andreas Dominik~Husch}

\affil[1]{
Luxembourg Centre for Systems Biomedicine
\protect\\
University of Luxembourg
\protect\\
7 Avenue des Hauts Fourneaux
\protect\\
4362, Esch-sur-Alzette
\protect\\
Luxembourg
\protect\\
beatriz.garcia@ext.uni.lu
\protect \\
\{jan.soelter, matias.bossa, andreas.husch\}@uni.lu
\vspace{0.5cm}
} 
\affil[2]{
National Dept. of Neurosurgery 
\protect\\
Centre Hospitalier de Luxembourg
\protect\\
4, Rue Ernest Barble
\protect\\
Luxembourg
}

\begin{document}
\maketitle
%
%
%


\begin{abstract}
Machine learning based methods for diagnosis and progression prediction of COVID-19 from imaging data have gained significant attention in the last months, in particular by the use of deep learning models. In this context hundreds of models where proposed with the majority of them trained on public datasets.
Data scarcity, mismatch between training and target population, group imbalance, and lack of documentation are important sources of bias, hindering the applicability of these models to real-world clinical practice.
Considering that datasets are an essential part of model building and evaluation, a deeper understanding of the current landscape is needed. 

This paper presents an overview of the currently public available COVID-19 chest X-ray datasets. Each dataset is briefly described and potential strength, limitations and interactions between datasets are identified. In particular, some key properties of current datasets that could be potential sources of bias, impairing models trained on them are pointed out.




These descriptions are useful for model building on those datasets, to choose the best dataset according the model goal, 
to take into account the specific limitations to avoid reporting overconfident benchmark results, and to discuss their impact on the generalisation capabilities in a specific clinical setting.    

\end{abstract}

\keywords{COVID-19 \and machine learning \and datasets \and X-Ray \and imaging \and overview \and bias \and confounding}


\section{Introduction}
\label{section:intro}

Just a bit more than half a year has passed since the novel corona virus SARS-CoV-2 gained world wide attention and eventually developed to the global COVID-19 pandemic~\cite{sohrabi2020world}. 
In course of the pandemic, diagnostics play a vital role in the management of cases and the distribution of potentially scarce resources, like hospital/ICU beds. 
Hence, there is an urgent necessity to create trustworthy tools for diagnosis and prognosis of the disease. While most of the people with COVID-19 infection do not develop pneumonia~\cite{cleverley2020role}, the identification of induced COVID-19 pneumonia cases is essential.
To this end, imaging studies such as X-ray and CT are employed. Chest X-rays (CXRs), while less sensitive that CT, are widely available, fast, non-invasive, and relatively cheap tests to diagnose and monitor COVID-19  induced pneumonia~\cite{aljondi2020diagnostic}.
Despite there is no single feature specifically associated with covid-19 induce pneumonia on a chest, its been associated with a combination of multifocal peripheral lung changes of ground glass opacity and/or consolidation, which are most commonly bilateral\cite{cleverley2020role} .


Machine learning, and especially deep learning methods, promise to assist the radiologist in coherent diagnosis and interpretation of images~\cite{choy2018current, mcbee2018deep}.
To this end, a remarkable amount of machine learning models has been proposed in this very short amount of time to tackle the problems of COVID-19 diagnosis, quantification and prognosis from X-Ray imaging \cite{shoeibi2020automated, islam2020review, ilyas2020detection}. 
But there is growing awareness in the community that the presence of different sources of bias significantly decreases overall generalisation of models, while model performances are overestimated in internal validation \cite{soneson2014batch,cohen2020limits, zech2018variable, maguolo2020critic}.

The most common sources of bias are unknown confounders and selections bias.
A confounder~\cite{greenland1999} is a variable that affects both, the potential predictor variable (the image in this case)
and the outcome (mainly diagnosis and radiological findings). If we are not aware of confounders, and do not control for them, we can erroneously conclude 
that a given image feature is a strong predictor of the outcome,
when in reality the association is spurious and it
does not hold any more in a different setting, where the confounder takes different values. 
Selection bias~\cite{Heckman79} happens when subjects
are not selected at random from the target population. If the image and the outcome affect the selection criteria 
a spurious association between image and outcome appears. As in the case of confounders, 
this association may not by present when the selection criteria are different and the model will misbehave. 
Other sources of bias~\cite{Castro2020Causality}, such as image noise, errors in outcome assignment or population prevalence imbalance could also lead to biased models. They can be mitigated in many cases, as far as 
the information to identify them is available.
Deep learning models present the additional inconvenient, with respect to traditional regression models, that it is not always easy to control for confounders, but this issue is outside the scope of this paper.

In order to avoid or at least detect bias, it is important that datasets and models are well documented. In a recent review and critical appraisal for prediction models for diagnosis and prognosis of COVID-19~\cite{wynants2020prediction} all evaluated models were rated at high risk of bias. Authors arrived
to the conclusion that they \say{do not recommend any of these reported prediction models for use in current practice}. The most common causes of risk of bias in diagnostic models based on medical imaging was: lack of information to assess selection bias (such as how controls were selected or which patients had imaging) 
and lack of clear report of image annotation procedures and of quality control.

To raise awareness of such problems, we aim to give a comprehensive overview on current publicly available chest X-ray datasets, identifying strengths as well as limitations, such as most evident potential sources of bias. 
To this end, we tried to reconstruct the story behind each dataset in section \ref{section:datasets}, describing the way images were obtained, which additional information is provided, and how the corresponding labels have been generated. Thus this overview 
provides a more in depth description of datasets than previous works~\cite{sohan2020so, shoeibi2020automated, islam2020review, ilyas2020detection}.    
Based on these findings, we discuss in section \ref{section:discussion} the specific risks of bias posed by these datasets. Finally, we give some concluding remarks in section \ref{section:remarks}.



\section{Publicly Available datasets} 
\label{section:datasets}
In this section we present a collection of the currently available public datasets for chest X-ray that present clinical manifestations of COVID-19 pneumonia. In addition, some non-COVID-19 Chest X-ray dataset were also included because they have been utilised in models for pre-training~\cite{cohen2020predicting} or to enrich datasets with control cases. 
Table \ref{tab:datasets} gives an overview on all presented datasets.   

\begin{table}[htbp]
\centering
\caption{Dataset collection for COVID-19 and non-COVID-19 Chest X-ray described in the paper. 2.2 Sections are COVID-19 oriented datasets, 2.3 are non-COVID19 oriented datasets and 2.4 correspond to compilation datasets}

%
\label{tab:datasets}
\resizebox{\textwidth}{!}{%
\setlength\extrarowheight{6 pt} 
\begin{threeparttable}
\begin{tabular}{cccc}
\toprule
\textbf{Section}       & \textbf{Dataset name}                                                     & \textbf{COVID-19}          &  \textbf{Origin}\\ 
\midrule
\ref{section:cohen}             & Cohen/IEEE 8032~\tnote{1} ~\cite{cohen2020covid}       & \cmark            & Global                                                                    \\ 
\ref{section:brixia}        & Brixia-score-COVID-19~\tnote{2} ~\cite{sig2020covid}  & \cmark          &\ re-annotated subset from Cohen/IEEE 8032 \\ 
\ref{section:GeneralBlockchain} & GeneralBlockchain  covid-19 ~\tnote{3}                & \cmark           &  re-annotated from Cohen/IEEE 8032  \\ 
\ref{section:figure1}     & agchung/Figure1  ~\tnote{4}                                                   & \cmark            & Global            \\
\ref{section:Hannover}    & ML Hannover ~\tnote{5}                                   & \cmark               & Germany                                                                   \\ 
\ref{section:actualmed}        & agchung/Actualmed  ~\tnote{6}                         & \cmark              & Spain                                                                     \\ 
\ref{section:Cancer_imagine}   & Cancer Imagine Archive   ~\tnote{7}                    & \cmark              & USA                                                                       \\ 
\ref{section:HM_hospitals}     & HM Hospitals      ~\tnote{8}                           & \cmark               & Spain                                                                     \\ 
\ref{section:BIMCV-PADCHEST}    & BIMCV-PADCHEST  ~\tnote{9}  ~~~~\cite{bustos2020padchest}  &                     & Spain                                                                     \\ 
\ref{section:BIMCV-PADCHEST-19} & \begin{tabular}[c]{@{}c@{}}BIMCV-COVID19\\ -PADCHEST\end{tabular} ~\tnote{10}  & & Spain                                                          \\ 
\ref{section:BIMCV-19}          & BIMCV-COVID19+     ~\tnote{10} ~~~\cite{vaya2020bimcv}   & \cmark             & Spain                                                            \\ 
\ref{section:CheXpert}          & CheXpert ~\tnote{11}   ~~~~\cite{irvin2019chexpert}    &                      & USA                                                                       \\ 
\ref{section:nih}              & ChestXray-NIH   ~\tnote{12} ~~~~\cite{wang2017chestx}              &                 & USA                                                                       \\ 
\ref{section:rsna}              & RSNA Pneunomia Kaggle    ~\tnote{13} ~~~~\cite{shih2019augmenting}       &      &  re-annotated subset from  ChestXray-NIH                                        \\ 
\ref{section:google}            & ChestXray-NIH Google ~\tnote{14} ~~~~\cite{majkowska2020chest}       &     &  re-annotated subset from  ChestXray-NIH                               \\ 
\ref{section:mog}                      & Montgomery   ~\tnote{15}         ~~~~\cite{jaeger2014two}             &       &        USA                                                       \\ 
\ref{section:shen}                            & Shenzhen    ~\tnote{16}   ~~~~\cite{jaeger2014two}        &    &     China                                            \\ 
\ref{section:UCSD}                           & UCSD-Guangzhou pediatric    ~~~~\tnote{17} ~~~~\cite{kermany2018identifying}             &               &           China           \\ 
\ref{section:MIMIC}                            & MIMIC-CXR-JPG v2.0.0    ~\tnote{18}    ~~~~\cite{johnson2019mimic}                                          &               &        Israel                            \\ 
\ref{section:Indiana}                            & Indiana University/OpenI
\tnote{19}   ~~\cite{demner2016preparing}                                &                &    USA                                                                       \\ 

 \ref{section:kagglecovid19}  & Kaggle COVID-19 radiography ~\tnote{20} ~~~\cite{chowdhury2020can}                                     & \cmark               &      Global                                                                      \\ 

 \ref{section:7labs}                           & V7 Darwin covid-19-chest-x-ray \tnote{21}                                 & \cmark              &        Global                                                                    \\ 

 \ref{section:COVIDx}                          & COVIDx~\tnote{22}  ~~~\cite{wang2020covidnet}                  &  \cmark               &            Global                                                               \\ 
\bottomrule
 \end{tabular}%
\begin{tablenotes}
\item[1] \url{https://github.com/ieee8023/covid-chestxray-dataset}
\item[2] \url{https://github.com/BrixIA/Brixia-score-COVID-19\#license-and-attribution}
\item[3] \url{https://github.com/GeneralBlockchain/covid-19-chest-xray-segmentations-dataset}
\item[4] \url{https://github.com/agchung/Figure1-COVID-chestxray-dataset}
\item[5] \url{https://github.com/ml-workgroup/covid-19-image-repository}
\item[6] \url{https://github.com/agchung/Actualmed-COVID-chestxray-dataset}
\item[7] \url{https://wiki.cancerimagingarchive.net/pages/viewpage.action?pageId=70226443}
\item[8] \url{https://www.hmhospitales.com/coronavirus/covid-data-save-lives/english-version}
\item[9] \url{https://bimcv.cipf.es/bimcv-projects/padchest}
\item[10] \url{https://github.com/BIMCV-CSUSP/BIMCV-COVID-19}
\item[11] \url{https://stanfordmlgroup.github.io/competitions/chexpert/}
\item[12] \url{https://nihcc.app.box.com/v/ChestXray-NIHCC}
\item[13] \url{https://www.kaggle.com/nih-chest-xrays/data}
\item[14] \url{https://www.kaggle.com/c/rsna-pneumonia-detection-challenge/overview/description}
\item[15] \url{http://openi.nlm.nih.gov/imgs/collections/NLM-MontgomeryCXRSet.zip}
\item[16] \url{http://openi.nlm.nih.gov/imgs/collections/ChinaSet_AllFiles.zip}
\item[17] \url{https://www.kaggle.com/paultimothymooney/chest-xray-pneumonia}
\item[18] \url{https://physionet.org/content/mimic-cxr-jpg/2.0.0}
\item[19] \url{https://openi.nlm.nih.gov/imgs/collections/NLMCXR_png.tgz}
\item[20] \url{https://www.kaggle.com/tawsifurrahman/covid19-radiography-database}
\item[21] \url{https://github.com/v7labs/covid-19-xray-dataset}
\item[22] \url{https://github.com/ieee8023/covid-chestxray-dataset}

\end{tablenotes}
\end{threeparttable}
}
\end{table}

\begin{figure}[htb]
\centering
\includegraphics[width=1.0\textwidth]{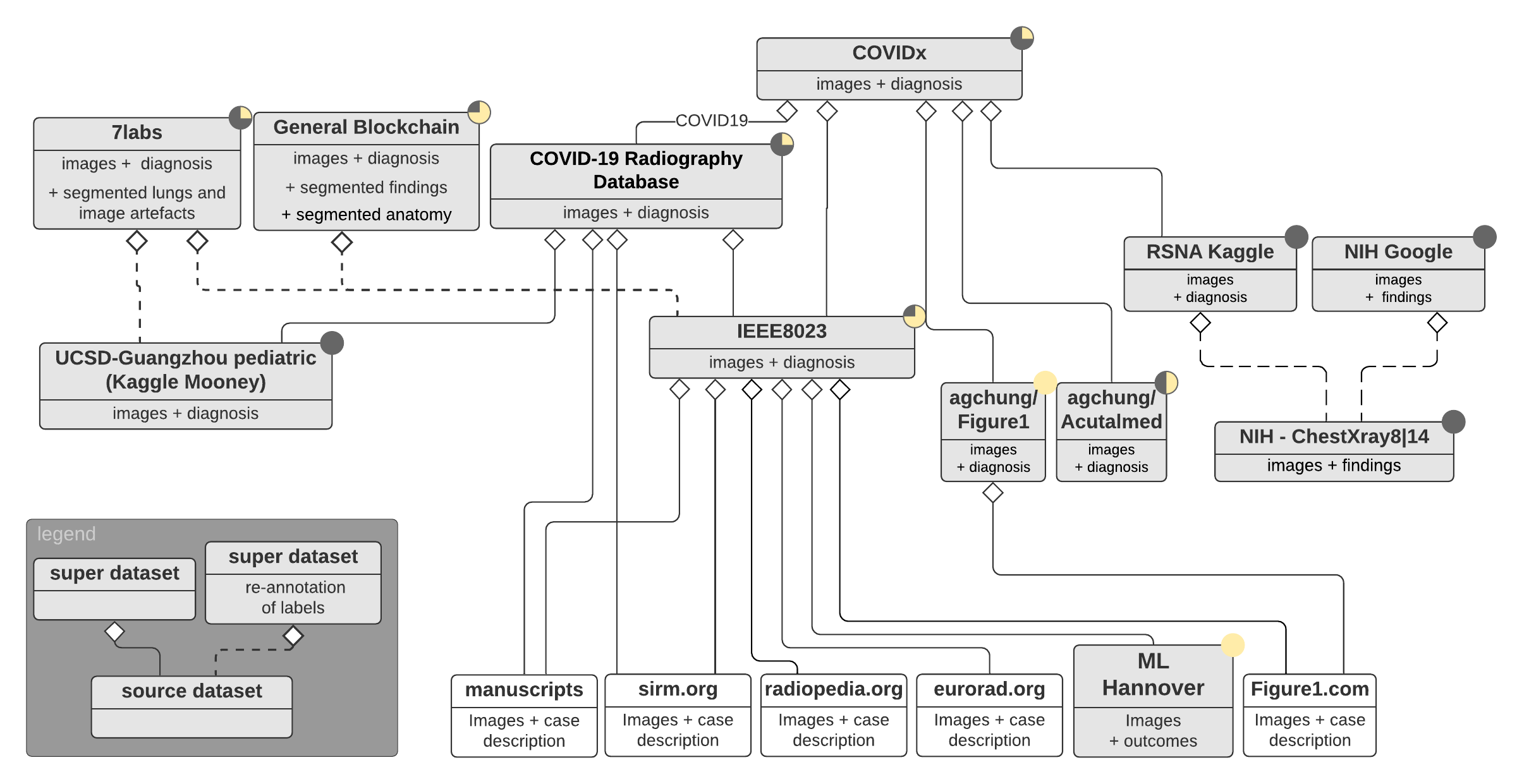}
\graphicspath{latex/images/bimcv_cf.png}
\label{fig:comparison} 
\caption{Overview on the relationship of interdependent datasets. Grey boxes represent datasets which can be downloaded as a single entity, whereas transparent boxes are general sources from where  information has to be extracted manually (or with automated scraping tools). Diamond shapes indicate that the attached dataset includes (parts) of the connected dataset. Dotted lines describe the case when images of a source dataset 
are included, but the labels are discarded and new labels have been obtained in a re-annotation process. The coloring of the circle in the upper right corner of each datasets gives the rough proportion of cases with radiological findings of COVID-19 (from full yellow for datasets with only COVID-19 to full gray for datasets without any COVID-19 case.)}
\end{figure}

\subsection{Radiological case reports of COVID-19} 
\label{section:websites}
Different radiological associations are making efforts to collect and provide images and reports on the internet to facilitate knowledge transfer of radiologist. Among those, \url{www.eurorad.org}, \url{www.sirm.org}, \url{www.radiopedia.org}, \url{www.figure1.com} and \url{www.bsti.org.uk} provide images which are usually accompanied by a radiological description and varying depth of clinical information. 
While these great initiatives allow for fast transfers of information during this emergency times, the extraction of both, image and meta-information, is not straightforward. Fortunately, several initiatives such as Cohen/IEEE 8032 dataset (\ref{section:cohen}) collect and present this information in a structured dataset format. 

\subsection{Datasets containing COVID-19 samples}

\subsubsection{Cohen/IEEE 8032}
\label{section:cohen}
The Cohen/IEEE 8032 dataset \cite{cohen2020covid} is a collection of cases 
\say{extracted from online publications, websites, or directly from the PDF}, including the aforementioned case sharing websites (\ref{section:websites}) . Additionally, it incorporates the ML Hannover dataset (\ref{section:Hannover}). It provides a well curated tabular view on the image meta-data containing the source document of the image, the imaging origin (hospital, city or country) and information about scanning view for most images, as well as gender, age 
and a variety of clinical features and outcomes for some cases. Each image is assigned a diagnosis of respiratory disease, with a strong focus on COVID-19 (currently 450 out of 650) and very little cases of no finding (20).Additionally the dataset contains global severity scores for ~100 images created in a post-hoc analysis of images according to a severity scheme~\cite{cohen2020predicting}.

\paragraph{Brixia-score-COVID-19}
\label{section:brixia}
This dataset is the result of a post-hoc annotation process two in which two radiologist labelled images of a private dataset and about 200 images of the pubic Cohen/IEEE 8032 dataset with  "BrixIA", i.e. localised severity scores \cite{sig2020covid}.

\paragraph{GeneralBlockchain COVID-19 segmentations}
\label{section:GeneralBlockchain}
This dataset
is a snapshot of the Cohen/IEEE 8032 (\ref{section:cohen}) in which two radiologist created, in a post-hoc annotation process, segmentation masks for anatomical parts, medical devices (e.g. tubes and probes) and radiological findings (Ground glass opacities and Consolidations).

\subsubsection{agchung/Figure1}
\label{section:figure1}
This dataset contains images of 48 patients extracted from the aforementioned case sharing website \url{www.figure1.com} (\ref{section:websites}). For some patients, it contains additional information about sex, temperature, pO2 saturation, the scanning view as well as some clinical notes. Most of images are assigned one of the labels COVID-19, pneumonia or no finding.

\subsubsection{ML Hannover}
\label{section:Hannover}
This dataset \cite{Winther2020} from the Institute for Diagnostic and Interventional Radiology (Hannover, Germany), contains ~240 chest-xray images (CR, DX) from 71 patients at different timepoints in the course of COVID-19. It contains metadata including scanning view (AP vs PA), patient master data, laboratory data and longitudinal information on admission, ICU-admission and death. 

\subsubsection{agchung/Actualmed}
\label{section:actualmed}
The Actualmed dataset 
contains images (CR, DX) of 215 patients. Each image is assigned a radiological diagnosis of "COVID19", "inconclusive" or "No finding". In addition it contains metadata of date and scanning view (AP vs PA). An analysis on the prevalence of COVID-19 findings within the different scanning views reveals a significantly higher prevalence of cases with COVID-19 cases in the AP scanning view.  

\begin{figure}[htb]
\centering
\includegraphics[scale=0.45]{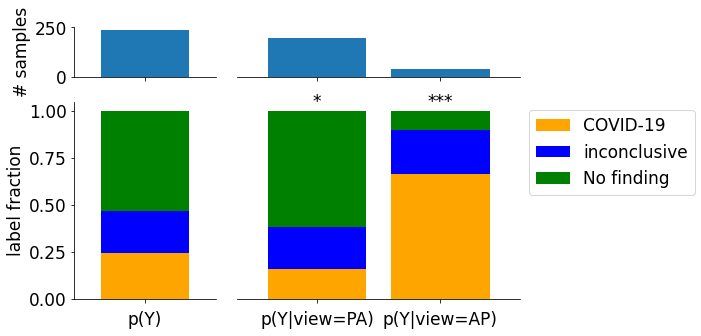}\
\graphicspath{latex/images/bias_actualmed.png}
\caption{Top panel: Number of samples in the agchung/Actualmed 
dataset, for the full dataset  (left) and for each scanning view (right). Bottom panel: fraction of associated labels in the full dataset (left) and conditioned on each scanning view (right). Stars mark the significance of the conditional distribution differing from the marginal distribution according to a $\chi^2$ test.}
\label{fig:biasact} 
\end{figure}

\subsubsection{Cancer Imagine Archive}
\label{section:Cancer_imagine}
This dataset was collected by the University of Arkansas for Medical Sciences (USA). It contains more than 200 chest X-ray studies for patients tested positive for COVID-19 including temporal data and key radiological findings. Moreover it provides extensive meta-information such as age, gender, race, several comorbidities and indicators of disease evolution such as ICU admission and mortality. Its focus lies on underrepresented rural populations in the USA. 

\subsubsection{HM Hospitals}
\label{section:HM_hospitals}
This dataset contains more that 5000 chest X-ray for COVID-19 (PCR positive or pending) in DICOM format, hence it contains information about gender, view, modality and others. Additional longitudinal clinical information includes admission times, treatment, vital signs, laboratory results and disease progression information such as ICU admissions, discharge or death. 
According to dataset owners, the intended use is to build prediction models of disease evolution, epidemiological models and information on the response to the various treatments. 
\subsubsection {BIMCV}
The Medical Imaging Databank of the Valencia Region (\b{B}anco digital de \b{I}magen \b{M}edica de la \b{C}omunidad \b{V}alenciana) provides both a general chest x-rays dataset from pre-COVID19 times and a dataset specially with COVID-19 confirmed cases.    

\paragraph{BIMCV-PADCHEST}
\label{section:BIMCV-PADCHEST}
The Padchest dataset \cite{bustos2020padchest} contains images (CR, DX, CT) of about 67000 patients together with their radiological reports from a single hospital (Hospital San Juan Hospital (Spain)) between 2009 and 2017.
It contains extensive DICOM meta-data including sex, age and scanner manufacturer. DICOM Scanning positions are given and grouped into 6 main classes (standard PA, standard L, AP vertical, AP horizontal, pediatric, and rib views) for images with position information available, otherwise a pre-trained CNN-model assigned automatically either one of the standard PA or L view. 

To facilitate model training, the authors mapped the radiological reports into a hierarchical set of categories, i.e.
\say{labels were extracted from the radiological reports, resulting in 22 differential diagnoses, 122 anatomic locations and 189 different radiological findings mapped onto standard Unified Medical Language System (UMLS)}, including the presence of medical devices. 
This mapping of the reports was carried out for 27\% of the dataset in a manual annotation process and completed for the rest of the data by a trained model. 

\paragraph{BIMCV-COVID19-PADCHEST}
\label{section:BIMCV-PADCHEST-19}
This is a subset of the BIMCV-PADCHEST dataset to four classes ("Control", "Infiltration without Pneumonia", "Pneumonia without Infiltration" and "Pneumonia with Infiltration"). It was created to facilitate (pre)-training of models related to COVID-19.

\paragraph{BIMCV-COVID19+}
\label{section:BIMCV-19}
This dataset contains images (CR, DX, CT) of 1311 patients together with their radiological reports collected between February and April 2020 in 11 different hospitals of the Valencia region \cite{vaya2020bimcv}. As within Padchest, 724 reports were manually mapped to the UMLS schema of the padchest dataset with two new categories (COVID19, COVID19-uncertain), and subsequently the remaining reports were mapped using a trained annotation model.
Additional to the radiological reporting, the dataset contains for each patient the result of one or more PCR-test obtained after the imaging (usually within 5 days), whereby each patient had at least one positive PCR test.
Furthermore in 10 images there are detailed ROI annotation for the radiological findings.

An analysis about prevalence of COVID-19 radiological findings with regard to the DICOM  "series description" shows a significant prevalence difference between the varying acquisition protocols. The least prevalence is observed in the "Erect PA" view, whereas the highest prevalence is present ind the "AP horizontal" and "Thorax AP" view.

\begin{figure}[htb]
\centering
\includegraphics[width=0.99\textwidth]{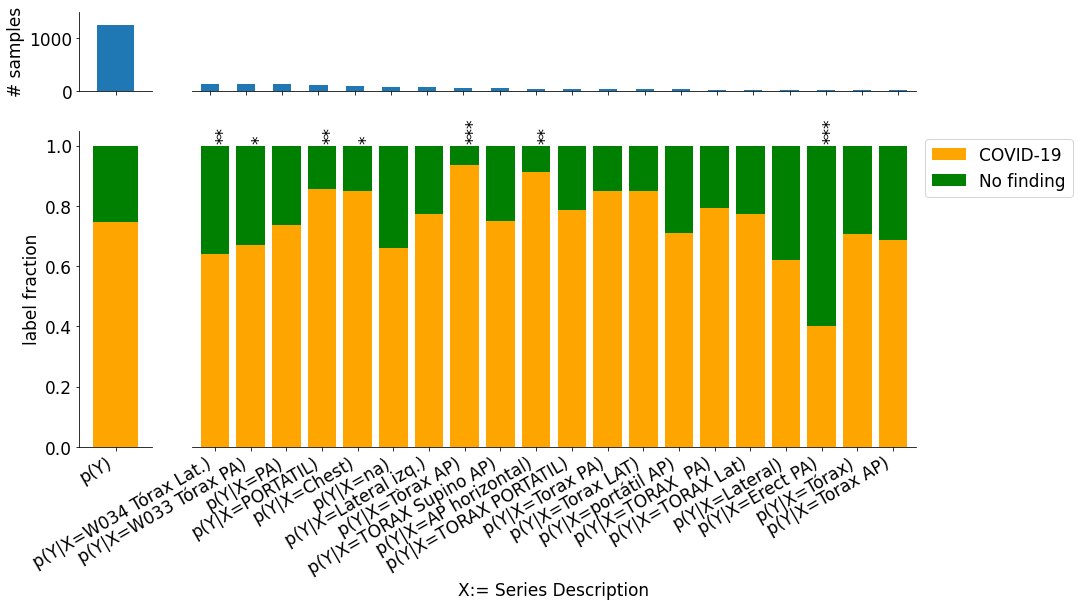}\
\caption{Top panel: Number of total samples in the full BIMCV-COVID19+ dataset (left) and in subsections partiioned by the DICOM metadata entry of Series Description (right). Bottom panel: Fraction of radiological findings in the full BIMCV-COVID19+ dataset (left) and when conditioned on the Series Description (right). Stars mark the significance of the conditional distribution differing from the marginal distribution according to a $\chi^2$ test.}
\label{fig:biasbimcv} 
\end{figure}

\subsection{Chest X-ray datasets without COVID-19 samples}
\subsubsection{CheXpert}
\label{section:CheXpert}
This large datasets \cite{irvin2019chexpert} released at the beginning of 2019 includes 224,316 chest radiography from 65,240 patients retrospectively collected from Stanford hospital. 
For fourteen classes of radiological findings associated with the most common respiratory diseases a rule-based processing of the reports generated labels of presence, absence, or uncertainty thereof. 
Additionally this dataset comprises a dedicated testset, with 500 images annotated by consensus of eight radiologist. 
 


\subsubsection{NIH}

\paragraph{ChestXray-NIH}
\label{section:nih}
This dataset~\cite{wang2017chestx} contains ~120k images of ~30k patients from the clinical PACS database at National Institutes of Health Clinical Center. In its original version, all images were labelled with originally eight (ChestXray8) and in the current version fourteen (ChestXray14) different findings from thoracic pathology. The labels have been extracted from the corresponding radiological reports by Natural Language Processing. The underlying reports are not publicly available, but the dataset contains meta-information about gender, age and view position. Additionally there are  annotated bounding boxes for all findings in about ~1000 images.

\paragraph{RSNA Pneunomia Kaggle}
\label{section:rsna}
This dataset \cite{shih2019augmenting} is a subset of 30k images from the ChestXray-NIH dataset with an enrichment of images with a Pneunomia related diagnosis. In a well defined annotation process 
a team of radiologist annotated areas of Lung Opacity 
with bounding boxes solely based on the information present in the image.

\paragraph{ChestXray-NIH Google}
\label{section:google}
This is another subset of ~18k images from the ChestXray-NIH dataset with extra labels~\cite{majkowska2020chest}. In another well designed annotation process radiologist assigned each image only with \say{access to the patient age and image view (PA/AP), but not to additional clinical or patient data} the presence/absence of four findings: pneumothorax, opacity, nodule/mass and fracture.

\subsubsection{Montgomery}
\label{section:mog}
This pre-COVID-19 pandemic dataset (2014) contains 138 frontal chest images collected in the department of health and human services, Mongomery Country, Maryland, USA. 
To each image, a short radiological report and a disease diagnosis is assigned (58 images with tuberculosis manifestations and 80 controls), as well as
a lung segmentation annotation automatically generated under the supervision of a radiologist using anatomical landmarks~\cite{candemir2013lung}.
The images themselves contain written markings of the scanning view (AP and PA) and there is additional metadata about gender and age.  
The intended used of the dataset is to boost of computer-aided diagnosis for pulmonary diseases with 
focus on TB~\cite{jaeger2014two}.

\subsubsection{Shenzhen}
\label{section:shen}
This dataset, released together with the Montgomery dataset contains images collected in collaboration the Shenzhen Hospital in China. It contains 662 frontal X-ray, of which 326 are normal and 336 contain TB manifestations. 
Additionally metadata includes sex, age and a short radiological description  \cite{jaeger2014two}. 

\subsubsection{UCSD-Guangzhou pediatric}
\label{section:UCSD}
This dataset contains more than 5000
chest X-ray images 
    from children (AP view) selected from retrospective cohorts of pediatric patients of 1-5 year old from Guangzhou Women and Children’s Medical Center, Guangzhou ~\cite{kermany2018identifying}. In an annotation process by two experts all images were assigned a diagnosis of viral/bacterial Pneumonia or Normal. No further meta-data is available.

\subsubsection{MIMIC-CXR-JPG v2.0.0}
\label{section:MIMIC}
This large pre-COVID-19 dataset comprise 377,110 chest x-rays associated with 227,827 de-identified imaging studies sourced from the Beth Israel Deaconess Medical Center. Images are provided with 14 labels derived from two natural language processing tools (NegBio and CheXpert) applied to the corresponding free-text radiology reports. \cite{johnson2019mimic}.


\subsubsection{Indiana University Chest X-rays/OpenI}
\label{section:Indiana}
This dataset from the Indiana University was created to provide a publicly available searchable database and  comprises of ~7500 Chest X-rays ~\cite{demner2016preparing}.
It includes view information, radiological reports including main findings and impressions. In a well documented post-hoc annotation process the reports have been mapped to localised findings, e.g. "Cicatrix/lung/base/left".
\subsection{Compilation datasets}



\subsubsection{Kaggle COVID-19 radiography database}
\label{section:kagglecovid19}
The Kaggle COVID-19 radiography database \cite{chowdhury2020can} is compiled from different datasets. It contains COVID19 cases form the aforementioned Cohen/IEEE 8032 (\ref{section:cohen}) as well as cases extracted from the sirm.org website and from 43 publications. Furthermore it incorporates the data of the UCSD-Guangzhou pediatric dataset (\ref{section:UCSD}) providing control cases and cases of viral pneumonia except COVID19. Thus the dataset contains the 3 labels COVID19, normal and viral pneumonia. The compiled dataset itself does not include any meta-information, except the image dataset source.

\subsubsection{V7 Darwin covid-19-chest-x-ray-dataset}
\label{section:7labs}
This is another dataset that is a compilation of the UCSD-Guangzhou pediatric dataset (\ref{section:UCSD}) for non-COVID-19 and the Cohen/IEEE 8032 for COVID-19 cases (\ref{section:cohen}). From a downstream annotation process it includes for all images extra lung segmentation masks and ignore masks marking radiologist's drawings (e.g. arrows) and medical devices overlapping the lungs.
\subsubsection{COVIDx}
\label{section:COVIDx}
The COVIDx dataset~\cite{wang2020covidnet} is compiled from different sub-datasets.
It contains the aforementioned Cohen/IEEE 8032 (\ref{section:cohen}), angchung/Actualmed (\ref{section:actualmed}) and angchung/Figure1 (\ref{section:figure1}) dataset and the COVID-19 cases of the  COVID-19 radioagraphy database (\ref{section:kagglecovid19}) as well as Pneunomia and Normal cases from the RSNA Kaggle Pneumonia dataset (\ref{section:rsna}). Thus each image is assigned one of the labels COVID-19, Pneunomia or Normal. As these sub-datasets contain common sources, the authors take care to remove duplication with the same source URL.
Nonetheless one should still acknowledge that there is the risk of case duplication due to incoherent source descriptions, e.g. it is possible to find occasional cases with suspicious similar metadata (Table \ref{tab:duplication})

\begin{table}[htb]
\centering
\caption{Example of two cases in the COVIDx dataset which are highly similar accroding to the available metadata}
\label{tab:duplication}
\begin{tabular}{ccccc}
\toprule
patientid & sex & age & temperature & notes 
\\
\midrule
150       & M   & 28  & 39.1 &
\makecell[lt]{
28M previously fit and well, 
\\ not on any regular medications, presented
\\ with a 6 day Hx of fever, non-productive
\\ cough and SOB for the last 4 days. His 
\\ symptoms started as sore throat and coryzal
\\ symptoms 8 days prior to his presentation
\\ and  he reported contact with a friend
\\  with similar symptomatology. [...]}
\\
COVID-00003a & M & 28 & 39.1 &
\makecell[lt]{
28M previously fit and well, 
\\ not on any regular medications, presented
\\ with a 6 day Hx of fever, non-productive
\\ cough and SOB for the last 4 days. His 
\\ symptoms started as sore throat and coryzal
\\ symptoms 8 days prior to his presentation
\\ and  he reported contact with a friend
\\  with similar symptomatology. [...]}
\\
\bottomrule
\end{tabular}
\end{table}


\section{Discussion}
\label{section:discussion}
Medical Imaging Models, especially Convolutional Neural Networks (CNNs), are known not only to learn underlying diagnostic features, but also to exploit confounding image information. For example, it was shown that the acquisition site, regarding both the hospital system and the specific department within a hospital, can be predicted with very high accuracy ($>99\%$)~\cite{zech2018variable}. If disease prevalence is associated with the acquisition site, as it is often the case, this can be a strong confounder.
Thus, in any composite dataset having separated sub-datasets for COVID-19 and control cases, the dataset is completely confounded with the group label and therefore it is difficult to isolate the disease effect from dataset effect, making learning almost impossible and posing a high risk of overestimating prediction performance. 
Indeed it has been observed that, by training on different COVID-19 and non-COVID-19 dataset combinations, the \say{deep model specialises not in recognising COVID features, but in  learning the common features [of the specific datasets]}~\cite{tartaglione2020unveiling}. Eventually, a CNN model is able to identify the source dataset with a high accuracy ($>90\%$) solely from the image border region containing no pathology information at all~\cite{maguolo2020critic}. 

Besides acquisition site, the demographic characteristics of populations can also be a strong confounder. Datasets that take cases from the UCSD-Guangzhou pediatric dataset (\ref{section:UCSD}) as non-COVID-19 examples (maximum age 5y) pose the risk that models will associate anatomical features of age with the diagnosis since, for example in the Cohen/IEEE 8032~(\ref{section:cohen}) dataset, the minimum age is 20y (mean 54y). 
If controlling for confounders is already difficult in deep learning models, normalising the images in such a wide and disjoint range of ages seems an impossible task.

But one not only has to be wary in composite datasets, also single source datasets are not free of potential confounders and other sources of bias. The classical example would be a different imaging protocol depending on the patient's health status. For example, the PA erect view is the preferred imaging view in general, but if the patient is not able to leave the bed it is much more common to do an AP view image. Indeed, in two exemplary datasets that contain labels of radiological findings, one can observe a significant change in prevalence of COVID-19 cases depending on the scanning view (see Figures \ref{fig:biasbimcv} and \ref{fig:biasact}).

Another confounding factor might be the presence of medical devices like ventilation equipment or ECG cables, which allows a model to associate images with patient treatment instead of disease status. For example,
for the NIH ChestXray14 dataset~(\ref{section:nih}), a critical evaluation has shown that \say{in the pneumothorax class, [...] 80\% of the positive cases have chest drains. In these examples, there were often no other features of pneumothorax}~\cite{oakden2020exploring}. Datasets which provide additional annotations on the presence of medical devices (e.g. "BIMCV", "7labs", "General Blockchain") facilitate a risk analysis on this confounding effect and also enable mitigation strategies in training.   

In general, one has to distinguish between labels which have been annotated by taking only the image itself into account, and labels which have been generated by a different source, i.e. from another diagnostic method like CT or PCR. Unfortunately, radilogical reports done in a in clinical routine are a mixture of both. Radiologist are often aware of the patients clinical context and this information is reflected in the reports, since they are done with the aim of communicating information between different doctors. For example, 
has been shown for the NIH ChestXray14 dataset~(\ref{section:nih}) that, in a substantial fraction of images, the associated finding extracted from the reports can not be confirmed by a post-hoc assessment of the the images~\cite{oakden2020exploring}.

Biases arise more easily when outcome labels and prediction model intended application is not clearly defined. If the model objective is to find radiological manifestation of the disease in the images that are not necessarily apparent to the radiologist naked eye, the labels should be the best possible diagnostic assessment obtained by any diagnostic test that doesn't include image information from same modality. For example, a perfectly righteous goal could be to determine whether a feature observed in CT, but not visible in XR, could be detected by subtle signals that ML models can identify. In contrast, if the goal is to reproduce radiological findings (for example, to save radiologist time) the label should be radiological annotations assessed by an independent clinician that has no information except for the image. Otherwise the risk of bias increases sensibly and the generalisation ability is compromised, because we can not really understand where the key information is coming from, what the model is learning, and what the possible sources of bias are.
In this sense, it's worth noting that a couple of datasets do provide such annotations solely derived from the images (RSNA Kaggle, NIH Google, General Blockchain, 7labs).

\section{Concluding Remarks}
\label{section:remarks}

We identify serious limitations in most, if not all, currently available datasets. It is urgently needed that more images from larger and better datasets are made publicly available. Dataset owners should make an effort to improve documentation about whole dataset building process to increase significantly the dataset value and the quality of models trained on them. For example: there should be a clear statement of dataset intended use, and explicit warning of common misuse cases; label definition and generating procedure should be reported in detail, so that other researchers can verify accuracy of label assignments and evaluate the utility and adequacy to the problem at hand; finally, datasets should contain cohort characteristics and subject selection criteria information, in order 
to evaluate the risk of selection bias and to check if the training and target population has similar characteristic.


Contrary to classical statistical models or standard machine learning methods, deep learning models are highly complex systems that may have several building steps. For reasons that range from avoiding overfitting to reducing memory and computation needs, some parts of the models are sometimes pre-trained, using specific dataset, and then kept fixed as other parts of the model are fintuned later. Quality standards of datasets used to pre-train these building blocks may not necessarily be as high as the ones for finetuning the final model, and some of these datasets that are deemed close to useless for training a serious medical diagnostic tool may be perfectly appropriate for pre-trainig steps. One only has to be cautious not including those subjects in the cross-validation loops of the final model.


Although dataset quality is the most important requirement for a medical diagnostic system to be reliable, other aspects of the model building are also prone to biases. Adherence to transparent practices, such as the TRIPOD (transparent reporting of a multivariable prediction model for individual prognosis or diagnosis) reporting guideline~\cite{TRIPOD}, and assessing risk of bias with PROBAST (Prediction model Risk Of Bias ASsessment Tool)~\cite{PROBAST}, would be a starting point. However extensions of these guidelines are required in order to be fully applicable to deep learning systems~\cite{wynants2020prediction}.

We hope our overview will help modellers to choose the appropriate dataset for their modelling needs while, at the same time, raise awareness on biases to look out for while training models. We also encourage everyone to validate models and report benchmarking results on a possibly small but very well curated external dataset, which is carefully selected to represent the real clinical use case as close as possible. Of course, in a best case scenario such gold standard test datasets would be publicly available to transparently foster progress in the field.

\section*{Acknowledgement}
This work was supported by the Luxembourg National Research Fund (FNR) Grant  COVID-19/2020-1/14702831/AICovIX/Husch. Beatriz Garcia Santa Cruz is supported by the FNR within the PARK-QC DTU (PRIDE17/12244779/PARK-QC).

\bibliographystyle{unsrt}  
\bibliography{references}  





\end{document}